%% file: sppl-probprog.tex
\documentclass[sigplan,10pt]{acmart}

\acmConference[PROBPROG'18]{The International Conference on Probabilistic Programming}{October 4--6, 2018}{Boston, MA}
\acmYear{*Equal Contribution, $\dag$Work at Oxford}
\acmISBN{} 
\acmDOI{} 
\startPage{1}

\setcopyright{none}

\usepackage{booktabs}   
\usepackage{subcaption} 

\usepackage{wrapfig}
\usepackage{paralist}
\usepackage{soul,color}
\usepackage{graphicx}
\graphicspath{{figures/}}
\usepackage{paralist}
\usepackage{caption}
\usepackage{subcaption}
\usepackage{setspace}
\usepackage{multicol}
\usepackage{adjustbox}
\usepackage{array}
\usepackage{multicol}
\usepackage{amsmath}
\usepackage{amsthm}
\usepackage{amssymb}

\usepackage{todonotes}

\definecolor{blue}{rgb}{0,0.3,0.7}
\definecolor{red}{rgb}{0.60,0.0,0.0}
\definecolor{purple}{rgb}{0.5,0,0.7}
\definecolor{cyan}{rgb}{0.0,0.6,0.5}
\definecolor{gray}{rgb}{0.4,0.4,0.4}
\input{anglican-color}
\lstdefinestyle{default}{language=Anglican,basicstyle=\ttfamily\small, columns=flexible, showstringspaces=false}
\newcommand{\ang}{\lstinline[mathescape,language=Anglican,basicstyle=\ttfamily\scriptsize]}

\newtheorem{theorem}{Theorem}

\newtheorem{defn}{Definition}

\usepackage{xcolor}
\usepackage{algorithm}
\usepackage{algpseudocode}

\begin{document}

\title{Hamiltonian Monte Carlo for Probabilistic Programs with Discontinuities}

\author{Yuan Zhou*}
\affiliation{\institution{University of Oxford}}
\author{Bradley J. Gram-Hansen*}
\affiliation{\institution{University of Oxford}}
\author{Tobias Kohn$\dag$}
\affiliation{\institution{University of Cambridge}} 
\author{Tom Rainforth}
\affiliation{\institution{University of Oxford}} 
\author{Hongseok Yang}
\affiliation{\institution{KAIST}}
\author{Frank Wood}
\affiliation{\institution{University of British Columbia}}
\renewcommand{\shortauthors}{Zhou and Gram-Hansen}

\begin{abstract}
Hamiltonian Monte Carlo (HMC) is arguably the dominant statistical inference algorithm used in most popular ``first-order differentiable''  Probabilistic Programming Languages (PPLs).  
However, the fact that HMC uses derivative information causes complications when the target distribution is non-differentiable with respect to one or more of the latent variables.

In this paper, we show how to use extensions to HMC to perform inference in probabilistic programs that contain discontinuities. 
To do this, we design a Simple first-order Probabilistic Programming Language (SPPL) that contains a sufficient set of language restrictions together with a compilation scheme.
This enables us to preserve both the statistical and syntactic interpretation of \texttt{if-else} statements in the probabilistic program, within the scope of first-order PPLs.
We also provide a corresponding mathematical formalism that ensures any joint density denoted in such a language has a suitably low measure of discontinuities. 

\end{abstract}


\maketitle

\vspace*{-1ex}
\section{Introduction}
Hamiltonian Monte Carlo~(HMC)~\cite{duane1987hybrid, neal2011mcmc} is an efficient Markov Chain Monte Carlo (MCMC)
algorithm that has been widely used for inference in a wide-range of probabilistic models~\cite{jacobs2016ovarian,svensson2017power,doyle2016robust}.  
Its superior performance arises from the advantageous
properties of the sample paths that it generates via Hamiltonian mechanics.
HMC, particularly the No-U-turn sampler (NUTS) \cite{hoffman2014no} variant,
is implemented in many Probabilistic Programming Systems (PPSs)\cite{fritz2017,tran2017deep, salvatier2016probabilistic, gelman2015stan}, and is the main inference engine of both PyMC3~\cite{salvatier2016probabilistic} and Stan~\cite{gelman2015stan,carpenter2015stan}.

One drawback of using HMC in probabilistic programming is that complications can arise when the target distribution is not differentiable with respect to one or more of its parameters.  
Developers often impose restrictions on the models that can be encoded to try and avoid these complications, such as preventing the use of discrete variables or only allowing them if they are directly marginalized out.

However, even these restrictions are not sufficient to guarantee the program is discontinuity-free because control flow special forms, such as \texttt{if-else} statements in PPLs, can also induce discontinuities. 

Though it turns out, perhaps surprisingly,  that HMC still constitutes a valid inference algorithm when the target is discontinuous or even if it has disjoint support, this can substantially reduce the acceptance rate leading to inefficient inference.

To mitigate this issue, several variants of HMC have been developed~\cite{afshar2015reflection, nishimura2017discontinuous} to perform inference on models with discontinuous densities.
However, existing systems, like Stan, are not able to support these variants as their compilation procedures do not unearth the required discontinuity information.
Creating a system that addresses this issue in an automated way is non-trivial and requires significant heavy lifting from the programming languages perspective.  

Therefore, in this extended abstract,  we define a carefully designed probabilistic programming language with an accompanying
denotational semantics and compiler.  
Together, these are able to both recover the discontinuity information required to use these HMC variants as inference engines and provide a framework amenable to the theoretical analysis required to demonstrate the correctness of the resulting engines.
To ensure that the measure of the set of discontinuities in the target density is of measure zero, we provide a mathematical formalism to compliment our language. This then provides us with a framework in which we can conservatively and correctly employ HMC and its variants.
We demonstrate this for the Discontinuous Hamiltonian Monte Carlo ~(DHMC) algorithm~\cite{nishimura2017discontinuous} and provide an example of our language being employed on models that have non-differentiable densities.
\section{A Simple PPL}
  
SPPL uses a Lisp style syntax,
like that of Church~\cite{goodman2012church}, Anglican~\cite{wood2014new} and Venture~\cite{venture2014}.
The syntax of expressions $e$ in our language is given as follows:
\vspace{-5pt}
\[
\begin{array}{@{}r@{}c@{}l@{}}
e & \,::=\, & 
x \mid c \mid\, (p\ e\,\ldots\,e) 
\,\mid\, (\texttt{\ang{if}}\ (< e\ 0)\ e\ e)
\,\mid\, (\texttt{\ang{let}}\ [x\ e]\ e)
\\
& \,\mid\, & (\texttt{\ang{sample}}\ (d\ e\, \ldots\, e))
\,\mid\, (\texttt{\ang{observe}}\ (d\ e\, \ldots\, e)\ c)
\vspace{-5pt}
\end{array}
\]
We use $x$ for a real-valued variable, $c$ for a real number, $p$ for an analytic primitive operation on
real, such as $+$ and $\exp{}$, and $d$ for the type of a distribution on
$\mathbb{R}$, such as \texttt{Normal}, that has a piece-wise smooth density under analytic partition. 

\begin{figure}[t]
	\centering
\includegraphics[width=\linewidth]{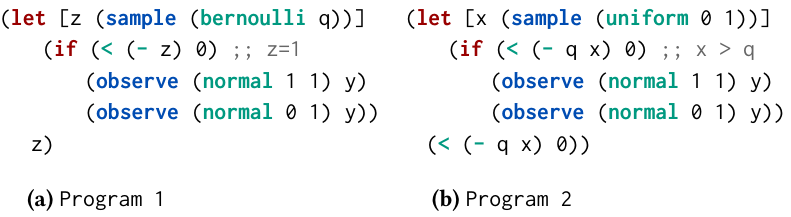}
	\vspace*{-4ex}
	\caption[]{A simple probabilistic program example written in Anglican (a) with Bernoulli distribution as a primitive and an equivalent one in SPPL (b) where the Bernoulli draw is constructed by a Uniform-draw and \texttt{if-else} statement.}
	\vspace*{-2ex}
	\label{fig:sppl}
\end{figure}

To be less technical, this is restricted on $d$
only allows continuous distribution as primitives. 
However, one can easily construct discrete distributions as was seen in Figure~\ref{fig:sppl},
where Program 1 (a) and 2 (b) define the joint density of the model respectively as following,
\begin{align*}
p_1(z, y) = &  \, p_{B}(z; q) 
\,  p_{\mathcal{N}}(y; 1,1)^{\mathbb{I}(z = 1)}
\, p_{\mathcal{N}}(y; 0,1)^{\mathbb{I}(z = 0)} \\
p_2(x, y) = & \,  p_{U}(x; 0,1)
\, p_{\mathcal{N}}(y; 1,1)^{\mathbb{I}(x >q)}
\, p_{\mathcal{N}}(y; 0,1)^{\mathbb{I}(x \leq q)}
\end{align*}

Note that there are no forms for applying general functions in this language and no recursion is possible at all.
As a result, all programs written in this language may only have a finite and fixed number of \ang{sample}  and \ang{observe} statements.
This means, among other things, that programs in this language can be compiled to
graphical models in which there are finite number of random variable vertices coming from every sample and observe statement.
For this reason we will mix our use of the terms graphical model and probabilistic program, or just program, because of their equivalence.

The primitives are, by design, restricted to be analytic. Analytic functions are abundant and most
primitive functions that we encounter in machine learning applications are analytic, and their composition  is also analytic. 

Intuitively, programs in SPPL have a joint density in the form as Definition~\ref{defn:piecewise-smooth-analytic} (See Appendix~\ref{sec:def1}). It can be understood as a collection of smooth functions in each partition, with no partition overlapping and the union of all partitions is the total space $\mathbb{R}^k$.
To evaluate the sum, therefore, we just need to evaluate
these products at $x$ one-by-one until we find one
that returns a non-zero value. Then, we have to compute
the function $\color{black}h_i$ corresponding to this product at the input $x$.

\begin{theorem}
	\label{thm:good-density-HMC}
	If the density $f : \mathbb{R}^{n} \to \mathbb{R}_+$
	has the form of Definition~\ref{defn:piecewise-smooth-analytic} and so is piecewise smooth under analytic partition,
	then there exists a (Borel) measurable subset $A \subseteq \mathbb{R}^{n}$ such
	that $f$ is differentiable outside of $A$ and the Lebesgue measure of $A$ is zero.
\end{theorem}

Together with Definition~\ref{defn:piecewise-smooth-analytic} and Theorem~\ref{thm:good-density-HMC},
we ensures that SPPL conforms to Lemma 1 and Theorems 2 and 3 of DHMC~\cite{nishimura2017discontinuous} and so, by design, all requirements for DHMC are trivially met as the density is a piecewise smooth function and all discontinuities are of measure zero.

\section{The Compilation Scheme}
Accompanying the language,
our compilation scheme is designed to establish variables which the density
is discontinuous with respect to and 
provide information for runtime checking on boundary crossing.
It works by automatically extracting \texttt{if} predicates and evaluating them with the corresponding random variables.
Each predicate is assigned a unique name and corresponding boolean. 
If the boolean changes value, 
it indicates the corresponding random variable has crossed the boundary during the update in inference at runtime.
The runtime checker will record this information and pass it to the inference engine.
Note that the runtime checking can only detect  
boundary crossing
instead of being able to calculate the analytical solution on where the boundaries are exactly.
We recognize the former part is sufficient for many specialized inference engines and leave the later part to be implemented as future work.

\section{Experiment}

\begin{figure}[t]
  \centering
  {\includegraphics[width=0.8\linewidth]{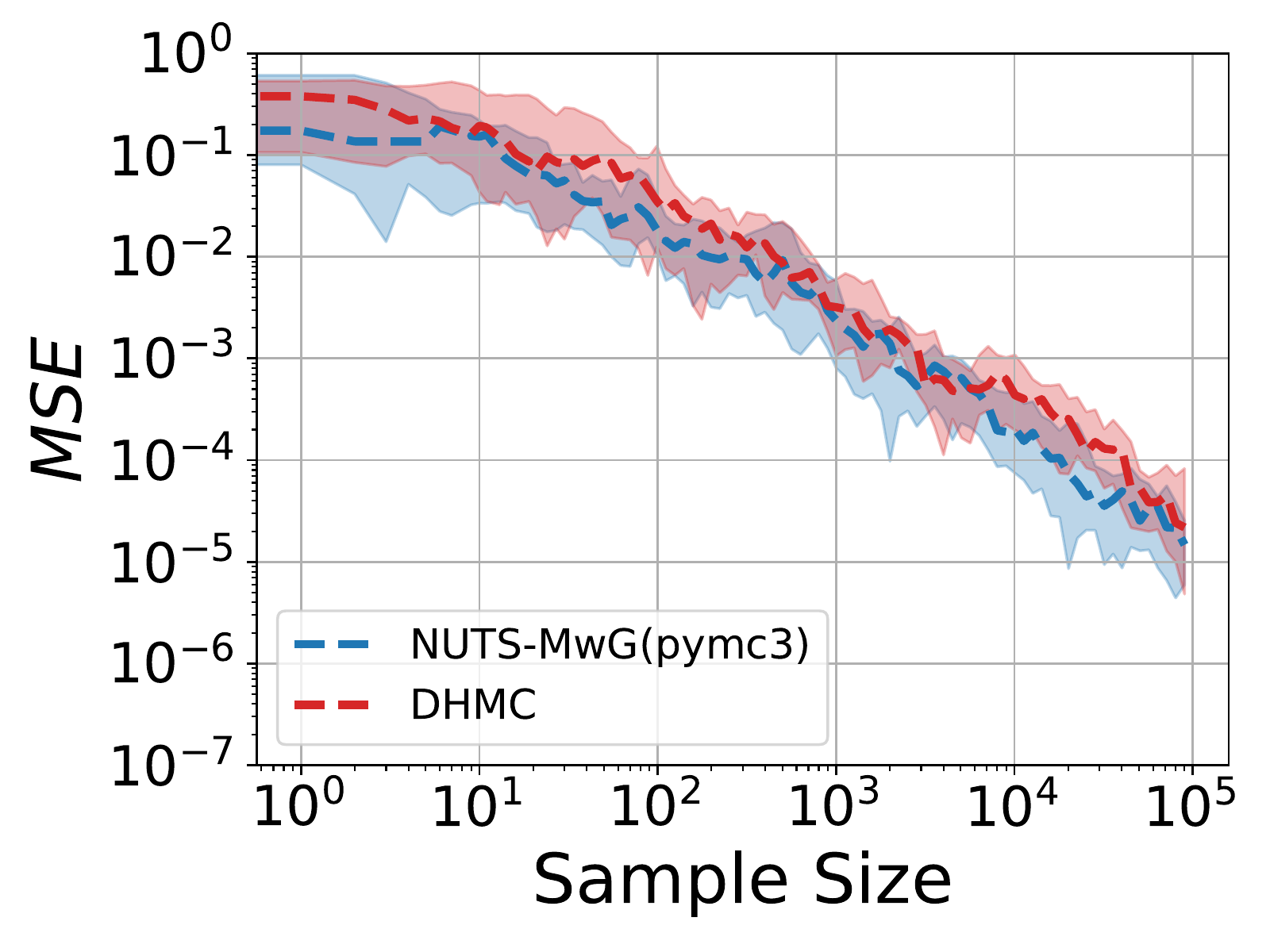}}
  \vspace*{-2ex}
	\caption{Mean Squared Error for the posterior estimates to true posterior of the cluster means $\mu_{1:2}$. We compare the results from DHMC and NUTS with Metropolis-within-Gibbs(MwG).
		The median (dashed lines) and $20\%/80\%$ confidence intervals(shaded) over $20$ independent runs are shown.}
  \vspace*{-2ex}
	\label{fig:gmm-s}
\end{figure}

We now consider a classic Gaussian mixture model (GMM),
where one is interested to estimate the mean of each cluster and the cluster assignment for each data.
The density of this model contains a mixture of continuous and discrete 
variables (See details of the model in Appendix~\ref{sec:appen-gmm}).

We compared the Mean Squared Error~(MSE) of the posterior estimates for the cluster means of both an unoptimized version of DHMC 
and PyMC3~\cite{salvatier2016probabilistic} optimized implementation of NUTS with Metropolis-within-Gibbs(MwG), with the same computation budget.
The results are shown in Figure~\ref{fig:gmm-s} as a function of number of
samples.
We take $100,000$ samples and discard $10,000$ for burn in. We calculate the $20\%/80\%$ confidence intervals over 20 independent runs
and find that both approaches perform consistently well. We find that our
unoptimized DHMC implementation, performs comparable to the
optimized NUTS with MwG approach.

\appendix
\section{Piecewise Smooth Function}
\label{sec:def1}
\begin{defn}
	\label{defn:piecewise-smooth-analytic}
	A function $f : \mathbb{R}^k \to \mathbb{R}$ is \emph{piecewise smooth
		under analytic partition} if it has the following form:
	\vspace{-5pt}
	\[\color{black}
	f(\color{black}x\color{black})\color{black}
	=
	\sum_{i = 1}^N
	\left(
	\prod_{j = 1}^{M_i}\mathbb{I}(\color{black}p_{i,j}\color{black}(\color{black}x\color{black}) \geq \color{black}0)\color{black}
	\cdot
	\prod_{l = 1}^{O_i}\mathbb{I}(\color{black}q_{i,l}\color{black}(\color{black}x\color{red}\color{black}) < \color{black}0)
	\cdot
	\color{black}h_i(\color{black}x\color{black})
	\right)
	\]
	\vspace{-2pt}
	where
	\vspace{-5pt}
	\begin{enumerate}
		\item the $\color{black}p_{i,j}, q_{i,l}\color{black} : \mathbb{R}^k \to \mathbb{R}$ are analytic;
		\item the $\color{black}h_i\color{black} : \mathbb{R}^k \to \mathbb{R}$ are smooth;
		\item $N$ is a non-negative integer or $\infty$;
		\item $M_i, O_i$ are non-negative integers; and
		\item the indicator functions
		\[
		\prod_{j = 1}^{M_i}\mathbb{I}(\color{black}p_{i,j}\color{black}(\color{black}x\color{black}) \geq \color{black}0)\color{black}
		\cdot
		\prod_{l = 1}^{O_i}\mathbb{I}(\color{black}q_{i,l}\color{black}(\color{black}x\color{red}\color{black}) < \color{black}0)
		\cdot
		\color{black}h_i(\color{black}x\color{black})
		\]
		\vspace{0pt}
		for the indices $i$ define a partition of $\mathbb{R}^k$ as,
		\[
		\left\{
		\left\{x \,{\in}\, \mathbb{R}^k
		\;\Big|\;
		\begin{array}{@{}l@{}}
		\color{black}p_{i,j}\color{black}(\color{black}x\color{black}) \,{\color{black}\geq}\, \color{black}0,\, \color{black}q_{i,l}(\color{black}x\color{black}) \,{\color{black}<}\, \color{black}0
		\mbox{ for all $j,l$}
		\end{array}\right\}
		\Big|\, 1 \,{\leq}\, i \,{\leq}\, N\right\}.
		\]
	\end{enumerate}
\end{defn}

\section{Proof of Theorem~\ref{thm:good-density-HMC}}

\begin{proof}
	Assume that $f$ is piecewise smooth under analytic partition. Thus,
	\begin{equation}
	\label{eqn:good-density-HMC1}
	f(x)
	=
	\sum_{i = 1}^N
	\prod_{j = 1}^{M_i} \mathbb{I}(p_{i,j}(x) \geq 0)
	\cdot
	\prod_{l = 1}^{O_i} \mathbb{I}(q_{i,l}(x) < 0)
	\cdot
	h_i(x)
	\end{equation}
	for some $N, M_i, O_i$ and $p_{i,j}, q_{i,l},h_i$ that satisfy the
	properties in Definition~\ref{defn:piecewise-smooth-analytic}.
	
	We use one well-known fact: the zero set $\{x \in \mathbb{R}^{n} \mid p(x) = 0\}$
	of an analytic function $p$ is the entire $\mathbb{R}^{n}$ or has zero Lebesgue measure~\cite{mityagin2015zero}.
	We apply the fact to each $p_{i,j}$ and deduce that
	the zero set of $p_{i,j}$ is $\mathbb{R}^{n}$ or has measure zero.
	Note that if the zero set of $p_{i,j}$ is the entire $\mathbb{R}^{n}$, the indicator
	function $[p_{i,j} \geq 0]$ becomes the constant-$1$ function, so that it can
	be omitted from the RHS of equation \eqref{eqn:good-density-HMC1}.
	In the rest of the proof, we assume that this simplification is already done so that
	the zero set of $p_{i,j}$ has measure zero for every $i,j$.
	
	For every $1 \leq i \leq N$, we decompose the $i$-th region
	\[
	R_i = \{x \mid p_{i,j} \geq 0\ \mbox{and}\ q_{i,l}(x) < 0\ \mbox{for all $j,l$}\}
	\]
	to
	\[
	\begin{array}{l}
	R'_i = \{x \mid p_{i,j} > 0\ \mbox{and}\ q_{i,l}(x) < 0\ \mbox{for all $j,l$}\}
	\\[0.5ex]
	\mbox{and}\ R''_i = R_i \setminus R'_i.
	\end{array}
	\]
	Note that $R'_i$ is open because the $p_{i,j}$ and $q_{i,l}$ are analytic and so continuous,
	both $\{r \in \mathbb{R} \mid r > 0\}$ and $\{r \in \mathbb{R} \mid r < 0\}$ are open,
	and the inverse images of open sets by continuous functions are open.
	This means that for each $x \in R'_i$, we can find an open ball at $x$ inside $R'_i$
	so that $f(x') = h_i(x')$
	for all $x'$ in the ball. Since $h_i$ is smooth, this implies that $f$ is differentiable
	at every $x \in R'_i$. For the other part $R''_i$, we notice that
	\[
	R''_i \subseteq \bigcup_{j = 1}^{M_i} \{x \mid p_{i,j}(x) = 0\}.
	\]
	The RHS of this equation is a finite union of measure-zero sets, 
	so it has measure zero. Thus, $R''_i$ also has measure zero as well.
	
	Since $\{R_i\}_{1\leq i \leq N}$ is a partition of $\mathbb{R}^{n}$, we have that
	\[
	\mathbb{R}^{n} = \bigcup_{i = 1}^N R'_i \cup \bigcup_{i=1}^N R''_i.
	\]
	The density $f$ is differentiable on the union of $R'_i$'s. Also, since
	the union of finitely or countably many measure-zero sets has measure zero,
	the union of $R''_i$'s has measure zero. Thus, we can set the set $A$ required
	in the theorem to be this second union.
\end{proof}

\section{Details of GMM}
The Gaussian Mixture Model~(GMM) can be defined as,
\label{sec:appen-gmm}
 \begin{align*}
\mu_{k} &\sim \mathcal{N}(\mu_0, \sigma_0), \; k = 1,\dots, K\\
z_{n} & \sim \text{Categorical}(p_0) , \; n = 1,\dots, N\\
y_n\, | z_n,\, & \mu_{z_n}  \sim \mathcal{N}(\mu_{z_n},\sigma_{z_n}), \; n = 1,\dots, N
\end{align*}
where $\mu_{1:K}, z_{1:N}$ are latent variables, $y_{1:N}$ are observed data 
with $ K$ as the number of clusters and $N$ the total number of data. 
Although Categorical distribution is not in our primitive, we can easily construct it in our language by the combination of uniform draws and nested \texttt{if} expressions.

For our experiment, we considered a simple case with 
$K = 2, \mu_0 = 0, \sigma_0 = 2, \sigma_{z_{1:N}} = 1$ and $p_0 = [0.5, 0.5]$, along with 
$y_{1:N} = [-2.0, -2.5,  -1.7,  -1.9,  -2.2, 1.5,  2.2,  3,  1.2, 2.8]$ as the
synthetic dataset. 
\bibliography{refs}

\end{document}

%% file: anglican-color.tex
\usepackage{listings}
\usepackage{color}
\usepackage{xcolor}
\definecolor{blue}{rgb}{0,0.3,0.7}
\definecolor{red}{rgb}{0.60,0.0,0.0}
\definecolor{purple}{rgb}{0.5,0,0.7}
\definecolor{cyan}{rgb}{0.0,0.6,0.5}
\definecolor{gray}{rgb}{0.4,0.4,0.4}

\lstdefinelanguage{scheme}
{sensitive, %
 alsoletter={:,-,+,*,?,/,!,>,<}, %
 morecomment=[l]{;}, %
}[comments]

\lstdefinelanguage{anglican}%
{
 morekeywords=[1]{},
 morekeywords=[2]{%
   def, def-, defn, defn-, defmacro, defmulti, defmethod, %
   defstruct, defonce, declare, definline, definterface, %
   defprotocol, defrecord, defstruct, deftype, defproject, ns, %
 }, %
 morekeywords=[3]{->, ->>, .., amap, and, areduce, as->, assert, binding, %
   bound-fn, case, comment, cond, cond->, cond->>, condp, declare, definline, %
   definterface, defmacro, defmethod, defmulti, defn, defn-, defonce, %
   defprotocol, defrecord, defstruct, deftype, delay, doseq, dosync, dotimes, %
   doto, extend-protocol, extend-type, fn, for, future, gen-class, %
   gen-interface, if, if-let, if-not, if-some, import, io!, lazy-cat, lazy-seq, let, %
   letfn, locking, loop, memfn, ns, or, proxy, proxy-super, pvalues, %
   recur, refer-clojure, reify, some->, some->>, sync, time, when, when-first, %
   when-let, when-not, when-some, while, with-bindings, with-in-str, %
   with-loading-context, with-local-vars, with-open, with-out-str, %
   with-precision, with-redefs, else}, %
  morekeywords=[4]{*, *', +, +', -, -', ->ArrayChunk, ->Vec, ->VecNode, %
    ->VecSeq, -cache-protocol-fn, -reset-methods, /, <, <=, =, ==, >, >=, %
    accessor, aclone, add-classpath, add-watch, agent, agent-error, %
    agent-errors, aget, alength, alias, all-ns, alter, alter-meta!, %
    alter-var-root, ancestors, apply, array-map, aset, aset-boolean, aset-byte, %
    aset-char, aset-double, aset-float, aset-int, aset-long, aset-short, assoc, %
    assoc!, assoc-in, associative?, atom, await, await-for, await1, bases, bean, %
    bigdec, bigint, biginteger, bit-and, bit-and-not, bit-clear, bit-flip, %
    bit-not, bit-or, bit-set, bit-shift-left, bit-shift-right, bit-test, %
    bit-xor, boolean, boolean-array, booleans, bound-fn*, bound?, butlast, byte, %
    byte-array, bytes, cast, char, char-array, char?, chars, chunk, %
    chunk-append, chunk-buffer, chunk-cons, chunk-first, chunk-next, chunk-rest, %
    chunked-seq?, class, class?, clear-agent-errors, clojure-version, coll?, %
    commute, comp, comparator, compare, compare-and-set!, compile, complement, %
    concat, conj, conj!, cons, constantly, construct-proxy, contains?, count, %
    counted?, create-ns, create-struct, cycle, dec, dec', decimal?, delay?, %
    deliver, denominator, deref, derive, descendants, destructure, disj, disj!, %
    dissoc, dissoc!, distinct, distinct?, doall, dorun, double, double-array, %
    doubles, drop, drop-last, drop-while, empty, empty?, ensure, %
    enumeration-seq, error-handler, error-mode, eval, even?, every-pred, every?, %
    ex-data, ex-info, extend, extenders, extends?, false?, ffirst, file-seq, %
    filter, filter-ns-publics, filterv, find, find-keyword, find-ns, %
    find-protocol-impl, find-protocol-method, find-var, first, flatten, float, %
    float-array, float?, floats, flush, fn?, fnext, fnil, force, format, %
    frequencies, future-call, future-cancel, future-cancelled?, future-done?, %
    future?, gensym, get, get-in, get-method, get-proxy-class, %
    get-thread-bindings, get-validator, group-by, hash, hash-combine, hash-map, %
    hash-ordered-coll, hash-set, hash-unordered-coll, identical?, identity, %
    ifn?, in-ns, inc, inc', init-proxy, instance?, int, int-array, integer?, %
    interleave, intern, interpose, into, into-array, ints, isa?, iterate, %
    iterator-seq, juxt, keep, keep-indexed, key, keys, keyword, keyword?, last, %
    line-seq, list, list*, list?, load, load-file, load-reader, load-string, %
    loaded-libs, long, long-array, longs, macroexpand, macroexpand-1, %
    make-array, make-hierarchy, map, map-indexed, map?, mapcat, mapv, max, %
    max-key, memoize, merge, merge-with, meta, method-sig, methods, min, %
    min-key, mix-collection-hash, mod, munge, name, namespace, namespace-munge, %
    neg?, newline, next, nfirst, nil?, nnext, not, not-any?, not-empty, %
    not-every?, not=, ns-aliases, ns-functions, ns-imports, ns-interns, %
    ns-macros, ns-map, ns-name, ns-publics, ns-refers, ns-resolve, ns-unalias, %
    ns-unmap, nth, nthnext, nthrest, num, number?, numerator, object-array, %
    odd?, parents, partial, partition, partition-all, partition-by, pcalls, %
    peek, persistent!, pmap, pop, pop!, pop-thread-bindings, pos?, pr, pr-str, %
    prefer-method, prefers, print, print-ctor, print-simple, print-str, printf, %
    println, println-str, prn, prn-str, promise, proxy-call-with-super, %
    proxy-mappings, proxy-name, push-thread-bindings, quot, rand, rand-int, %
    rand-nth, range, ratio?, rational?, rationalize, re-find, re-groups, %
    re-matcher, re-matches, re-pattern, re-seq, read, read-line, read-string, %
    realized?, record?, reduce, reduce-kv, reduced, reduced?, reductions, ref, %
    ref-history-count, ref-max-history, ref-min-history, ref-set, refer, %
    release-pending-sends, rem, remove, remove-all-methods, remove-method, %
    remove-ns, remove-watch, repeat, repeatedly, replace, replicate, require, %
    reset!, reset-meta!, resolve, rest, restart-agent, resultset-seq, reverse, %
    reversible?, rseq, rsubseq, satisfies?, second, select-keys, send, send-off, %
    send-via, seq, seq?, seque, sequence, sequential?, set, %
    set-agent-send-executor!, set-agent-send-off-executor!, set-error-handler!, %
    set-error-mode!, set-validator!, set?, short, short-array, shorts, shuffle, %
    shutdown-agents, slurp, some, some-fn, some?, sort, sort-by, sorted-map, %
    sorted-map-by, sorted-set, sorted-set-by, sorted?, special-symbol?, spit, %
    split-at, split-with, str, string?, struct, struct-map, subs, subseq, %
    subvec, supers, swap!, symbol, symbol?, take, take-last, take-nth, %
    take-while, test, the-ns, thread-bound?, to-array, to-array-2d, trampoline, %
    transient, tree-seq, true?, type, unchecked-add, unchecked-add-int, %
    unchecked-byte, unchecked-char, unchecked-dec, unchecked-dec-int, %
    unchecked-divide-int, unchecked-double, unchecked-float, unchecked-inc, %
    unchecked-inc-int, unchecked-int, unchecked-long, unchecked-multiply, %
    unchecked-multiply-int, unchecked-negate, unchecked-negate-int, %
    unchecked-remainder-int, unchecked-short, unchecked-subtract, %
    unchecked-subtract-int, underive, unsigned-bit-shift-right, update-in, %
    update-proxy, use, val, vals, var-get, var-set, var?, vary-meta, vec, %
    vector, vector-of, vector?, with-bindings*, with-meta, with-redefs-fn, %
    xml-seq, zero?, zipmap}, %
  morekeywords=[5]{def-cps-fn, defanglican, defm, defquery, defun, defproc, defdist}, %
  morekeywords=[6]{cps-fn, fm, lambda, query, with-primitive-procedures}, %
  morekeywords=[7]{%
    doquery, %
    conditional, %
    collect-by, equalize, exec, infer, log-marginal, print-predicts, %
    rand, rand-int, rand-nth, rand-roulette, stripdown, warmup, %
    ->CRP-process, ->DP-process, ->GP-process, %
    ->bernoulli-distribution, ->beta-distribution, ->binomial-distribution, %
    ->categorical-crp-distribution, ->categorical-distribution, %
    ->categorical-dp-distribution, ->chi-squared-distribution, %
    ->dirichlet-distribution, ->discrete-distribution, %
    ->exponential-distribution, ->flip-distribution, ->gamma-distribution, %
    ->mvn-distribution, ->normal-distribution, ->poisson-distribution, %
    ->sample, ->observe, sample*, observe*, %
    ->uniform-continuous-distribution, ->uniform-discrete-distribution, %
    ->wishart-distribution, CRP, DP, GP, abs, absorb, acos, asin, atan, %
    bernoulli, beta, binomial, categorical, categorical-crp, categorical-dp, %
    cbrt, ceil, chi-squared, cos, cosh, cov, dirichlet, discrete, exp, %
    exponential, flip, floor, gamma, gen-matrix, log, log-gamma-fn, %
    log-mv-gamma-fn, log-sum-exp, map->CRP-process, map->DP-process, %
    map->GP-process, map->bernoulli-distribution, map->beta-distribution, %
    map->binomial-distribution, map->categorical-crp-distribution, %
    map->categorical-distribution, map->categorical-dp-distribution, %
    map->chi-squared-distribution, map->dirichlet-distribution, %
    map->discrete-distribution, map->exponential-distribution, %
    map->flip-distribution, map->gamma-distribution, map->mvn-distribution, %
    map->normal-distribution, map->poisson-distribution, %
    map->uniform-continuous-distribution, map->uniform-discrete-distribution, %
    map->wishart-distribution, mvn, normal, poisson, pow, produce, %
    rint, round, signum, sin, sinh, sqrt, tag, tan, tanh, transform-sample, %
    uniform-continuous, uniform-discrete, wishart, uniform, %
    add-log-weight, add-predict, clear-predicts, get-log-weight, %
    get-mem, get-predicts, in-mem?, set-log-weight, set-mem, %
  }, %
  morekeywords=[8]{factor, mem, observe, predict, retrieve, sample, store}, %
  sensitive, %
  alsoletter={:,-,+,*,?,/,!,>,<,.}, %
  morecomment=[l][\color{gray}]{;}, %
  morestring=[b]", %
}[keywords,comments,strings]
 

%% file: sppl-probprog.bbl

\begin{thebibliography}{17}


\ifx \showCODEN    \undefined \def \showCODEN     #1{\unskip}     \fi
\ifx \showDOI      \undefined \def \showDOI       #1{#1}\fi
\ifx \showISBNx    \undefined \def \showISBNx     #1{\unskip}     \fi
\ifx \showISBNxiii \undefined \def \showISBNxiii  #1{\unskip}     \fi
\ifx \showISSN     \undefined \def \showISSN      #1{\unskip}     \fi
\ifx \showLCCN     \undefined \def \showLCCN      #1{\unskip}     \fi
\ifx \shownote     \undefined \def \shownote      #1{#1}          \fi
\ifx \showarticletitle \undefined \def \showarticletitle #1{#1}   \fi
\ifx \showURL      \undefined \def \showURL       {\relax}        \fi
\providecommand\bibfield[2]{#2}
\providecommand\bibinfo[2]{#2}
\providecommand\natexlab[1]{#1}
\providecommand\showeprint[2][]{arXiv:#2}

\bibitem[\protect\citeauthoryear{Afshar and Domke}{Afshar and Domke}{2015}]%
        {afshar2015reflection}
\bibfield{author}{\bibinfo{person}{Hadi~Mohasel Afshar} {and}
  \bibinfo{person}{Justin Domke}.} \bibinfo{year}{2015}\natexlab{}.
\newblock \showarticletitle{{Reflection, Refraction, and Hamiltonian Monte
  Carlo}}. In \bibinfo{booktitle}{\emph{Advances in Neural Information
  Processing Systems}}. \bibinfo{pages}{3007--3015}.
\newblock


\bibitem[\protect\citeauthoryear{Carpenter, Hoffman, Brubaker, Lee, Li, and
  Betancourt}{Carpenter et~al\mbox{.}}{2015}]%
        {carpenter2015stan}
\bibfield{author}{\bibinfo{person}{Bob Carpenter}, \bibinfo{person}{Matthew~D
  Hoffman}, \bibinfo{person}{Marcus Brubaker}, \bibinfo{person}{Daniel Lee},
  \bibinfo{person}{Peter Li}, {and} \bibinfo{person}{Michael Betancourt}.}
  \bibinfo{year}{2015}\natexlab{}.
\newblock \showarticletitle{The Stan Math Library: Reverse-mode Automatic
  Differentiation in C++}.
\newblock \bibinfo{journal}{\emph{arXiv preprint arXiv:1509.07164}}
  (\bibinfo{year}{2015}).
\newblock


\bibitem[\protect\citeauthoryear{Doyle, Yurovsky, and Frank}{Doyle
  et~al\mbox{.}}{2016}]%
        {doyle2016robust}
\bibfield{author}{\bibinfo{person}{Gabriel Doyle}, \bibinfo{person}{Dan
  Yurovsky}, {and} \bibinfo{person}{Michael~C Frank}.}
  \bibinfo{year}{2016}\natexlab{}.
\newblock \showarticletitle{A Robust Framework for Estimating Linguistic
  Alignment in Twitter Conversations}. In \bibinfo{booktitle}{\emph{Proceedings
  of the 25th international conference on world wide web}}. International World
  Wide Web Conferences Steering Committee, \bibinfo{pages}{637--648}.
\newblock


\bibitem[\protect\citeauthoryear{Duane, Kennedy, Pendleton, and Roweth}{Duane
  et~al\mbox{.}}{1987}]%
        {duane1987hybrid}
\bibfield{author}{\bibinfo{person}{Simon Duane}, \bibinfo{person}{Anthony~D
  Kennedy}, \bibinfo{person}{Brian~J Pendleton}, {and} \bibinfo{person}{Duncan
  Roweth}.} \bibinfo{year}{1987}\natexlab{}.
\newblock \showarticletitle{{Hybrid Monte Carlo}}.
\newblock \bibinfo{journal}{\emph{Physics letters B}} (\bibinfo{year}{1987}).
\newblock


\bibitem[\protect\citeauthoryear{Eli, Chen, Jankowiak, Karaletsos, Obermeyer,
  Pradhan, Singh, Szerlip, and Goodman}{Eli et~al\mbox{.}}{2017}]%
        {fritz2017}
\bibfield{author}{\bibinfo{person}{Bingham Eli}, \bibinfo{person}{Jonathan~P
  Chen}, \bibinfo{person}{Martin Jankowiak}, \bibinfo{person}{Theofanis
  Karaletsos}, \bibinfo{person}{Fritz Obermeyer}, \bibinfo{person}{Neeraj
  Pradhan}, \bibinfo{person}{Rohit Singh}, \bibinfo{person}{Paul Szerlip},
  {and} \bibinfo{person}{Noah Goodman}.} \bibinfo{year}{2017}\natexlab{}.
\newblock \bibinfo{title}{Pyro: Deep Probabilistic Programming}.
\newblock \bibinfo{howpublished}{\url{https://github.com/uber/pyro}}.
\newblock


\bibitem[\protect\citeauthoryear{Gelman, Lee, and Guo}{Gelman
  et~al\mbox{.}}{2015}]%
        {gelman2015stan}
\bibfield{author}{\bibinfo{person}{Andrew Gelman}, \bibinfo{person}{Daniel
  Lee}, {and} \bibinfo{person}{Jiqiang Guo}.} \bibinfo{year}{2015}\natexlab{}.
\newblock \showarticletitle{{Stan: A Probabilistic Programming Language for
  Bayesian Inference and Optimization}}.
\newblock \bibinfo{journal}{\emph{Journal of Educational and Behavioral
  Statistics}} \bibinfo{volume}{40}, \bibinfo{number}{5}
  (\bibinfo{year}{2015}), \bibinfo{pages}{530--543}.
\newblock


\bibitem[\protect\citeauthoryear{Goodman, Mansinghka, Roy, Bonawitz, and
  Tenenbaum}{Goodman et~al\mbox{.}}{2008}]%
        {goodman2012church}
\bibfield{author}{\bibinfo{person}{Noah~D. Goodman}, \bibinfo{person}{Vikash~K.
  Mansinghka}, \bibinfo{person}{Daniel~M. Roy}, \bibinfo{person}{Keith
  Bonawitz}, {and} \bibinfo{person}{Joshua~B. Tenenbaum}.}
  \bibinfo{year}{2008}\natexlab{}.
\newblock \showarticletitle{{Church: A Language for Generative Models}}. In
  \bibinfo{booktitle}{\emph{In UAI}}. \bibinfo{pages}{220--229}.
\newblock


\bibitem[\protect\citeauthoryear{Hoffman and Gelman}{Hoffman and
  Gelman}{2014}]%
        {hoffman2014no}
\bibfield{author}{\bibinfo{person}{Matthew~D Hoffman} {and}
  \bibinfo{person}{Andrew Gelman}.} \bibinfo{year}{2014}\natexlab{}.
\newblock \showarticletitle{{The No-U-turn Sampler: Adaptively Setting Path
  Lengths in Hamiltonian Monte Carlo.}}
\newblock \bibinfo{journal}{\emph{Journal of Machine Learning Research}}
  \bibinfo{volume}{15}, \bibinfo{number}{1} (\bibinfo{year}{2014}),
  \bibinfo{pages}{1593--1623}.
\newblock


\bibitem[\protect\citeauthoryear{Jacobs, Menon, Ryan, Gentry-Maharaj, Burnell,
  Kalsi, Amso, Apostolidou, Benjamin, Cruickshank, et~al\mbox{.}}{Jacobs
  et~al\mbox{.}}{2016}]%
        {jacobs2016ovarian}
\bibfield{author}{\bibinfo{person}{Ian~J Jacobs}, \bibinfo{person}{Usha Menon},
  \bibinfo{person}{Andy Ryan}, \bibinfo{person}{Aleksandra Gentry-Maharaj},
  \bibinfo{person}{Matthew Burnell}, \bibinfo{person}{Jatinderpal~K Kalsi},
  \bibinfo{person}{Nazar~N Amso}, \bibinfo{person}{Sophia Apostolidou},
  \bibinfo{person}{Elizabeth Benjamin}, \bibinfo{person}{Derek Cruickshank},
  {et~al\mbox{.}}} \bibinfo{year}{2016}\natexlab{}.
\newblock \showarticletitle{Ovarian Cancer Screening and Mortality in the UK
  Collaborative Trial of Ovarian Cancer Screening (UKCTOCS): A Randomised
  Controlled Trial}.
\newblock \bibinfo{journal}{\emph{The Lancet}} \bibinfo{volume}{387},
  \bibinfo{number}{10022} (\bibinfo{year}{2016}), \bibinfo{pages}{945--956}.
\newblock


\bibitem[\protect\citeauthoryear{Mansinghka, Selsam, and Perov}{Mansinghka
  et~al\mbox{.}}{2014}]%
        {venture2014}
\bibfield{author}{\bibinfo{person}{Vikash Mansinghka}, \bibinfo{person}{Daniel
  Selsam}, {and} \bibinfo{person}{Yura Perov}.}
  \bibinfo{year}{2014}\natexlab{}.
\newblock \showarticletitle{Venture: A Higher-order Probabilistic Programming
  Platform with Programmable Inference}.
\newblock \bibinfo{journal}{\emph{arXiv preprint arXiv:1404.0099}}
  (\bibinfo{year}{2014}).
\newblock


\bibitem[\protect\citeauthoryear{Mityagin}{Mityagin}{2015}]%
        {mityagin2015zero}
\bibfield{author}{\bibinfo{person}{Boris Mityagin}.}
  \bibinfo{year}{2015}\natexlab{}.
\newblock \showarticletitle{{The Zero Set of a Real Analytic Function}}.
\newblock \bibinfo{journal}{\emph{arXiv preprint arXiv:1512.07276}}
  (\bibinfo{year}{2015}).
\newblock


\bibitem[\protect\citeauthoryear{Neal}{Neal}{2011}]%
        {neal2011mcmc}
\bibfield{author}{\bibinfo{person}{Radford~M Neal}.}
  \bibinfo{year}{2011}\natexlab{}.
\newblock \showarticletitle{{MCMC Using Hamiltonian dynamics}}.
\newblock \bibinfo{journal}{\emph{Handbook of Markov Chain Monte Carlo}}
  (\bibinfo{year}{2011}).
\newblock


\bibitem[\protect\citeauthoryear{Nishimura, Dunson, and Lu}{Nishimura
  et~al\mbox{.}}{2017}]%
        {nishimura2017discontinuous}
\bibfield{author}{\bibinfo{person}{Akihiko Nishimura}, \bibinfo{person}{David
  Dunson}, {and} \bibinfo{person}{Jianfeng Lu}.}
  \bibinfo{year}{2017}\natexlab{}.
\newblock \showarticletitle{{Discontinuous Hamiltonian Monte Carlo for Sampling
  Discrete Parameters}}.
\newblock \bibinfo{journal}{\emph{arXiv preprint arXiv:1705.08510}}
  (\bibinfo{year}{2017}).
\newblock


\bibitem[\protect\citeauthoryear{Salvatier, Wiecki, and Fonnesbeck}{Salvatier
  et~al\mbox{.}}{2016}]%
        {salvatier2016probabilistic}
\bibfield{author}{\bibinfo{person}{John Salvatier}, \bibinfo{person}{Thomas~V
  Wiecki}, {and} \bibinfo{person}{Christopher Fonnesbeck}.}
  \bibinfo{year}{2016}\natexlab{}.
\newblock \showarticletitle{{Probabilistic Programming in Python Using PyMC3}}.
\newblock \bibinfo{journal}{\emph{PeerJ Computer Science}}  \bibinfo{volume}{2}
  (\bibinfo{year}{2016}), \bibinfo{pages}{e55}.
\newblock


\bibitem[\protect\citeauthoryear{Svensson, Natarajan, Ly, Miragaia, Labalette,
  Macaulay, Cvejic, and Teichmann}{Svensson et~al\mbox{.}}{2017}]%
        {svensson2017power}
\bibfield{author}{\bibinfo{person}{Valentine Svensson},
  \bibinfo{person}{Kedar~Nath Natarajan}, \bibinfo{person}{Lam-Ha Ly},
  \bibinfo{person}{Ricardo~J Miragaia}, \bibinfo{person}{Charlotte Labalette},
  \bibinfo{person}{Iain~C Macaulay}, \bibinfo{person}{Ana Cvejic}, {and}
  \bibinfo{person}{Sarah~A Teichmann}.} \bibinfo{year}{2017}\natexlab{}.
\newblock \showarticletitle{Power Analysis of Single-cell RNA-sequencing
  Experiments}.
\newblock \bibinfo{journal}{\emph{Nature methods}} \bibinfo{volume}{14},
  \bibinfo{number}{4} (\bibinfo{year}{2017}), \bibinfo{pages}{381}.
\newblock


\bibitem[\protect\citeauthoryear{Tran, Hoffman, Saurous, Brevdo, Murphy, and
  Blei}{Tran et~al\mbox{.}}{2017}]%
        {tran2017deep}
\bibfield{author}{\bibinfo{person}{Dustin Tran}, \bibinfo{person}{Matthew~D
  Hoffman}, \bibinfo{person}{Rif~A Saurous}, \bibinfo{person}{Eugene Brevdo},
  \bibinfo{person}{Kevin Murphy}, {and} \bibinfo{person}{David~M Blei}.}
  \bibinfo{year}{2017}\natexlab{}.
\newblock \showarticletitle{Deep Probabilistic Programming}.
\newblock \bibinfo{journal}{\emph{arXiv preprint arXiv:1701.03757}}
  (\bibinfo{year}{2017}).
\newblock


\bibitem[\protect\citeauthoryear{Wood, Meent, and Mansinghka}{Wood
  et~al\mbox{.}}{2014}]%
        {wood2014new}
\bibfield{author}{\bibinfo{person}{Frank Wood}, \bibinfo{person}{Jan~Willem
  Meent}, {and} \bibinfo{person}{Vikash Mansinghka}.}
  \bibinfo{year}{2014}\natexlab{}.
\newblock \showarticletitle{{A New Approach to Probabilistic Programming
  Inference}}. In \bibinfo{booktitle}{\emph{Artificial Intelligence and
  Statistics}}.
\newblock


\end{thebibliography}
